\documentstyle[preprint,revtex]{aps}
\begin{document}
\draft
\begin{title}
Non-Analytic Contributions to the Self-Energy and
the Thermodynamics of Two-Dimensional Fermi Liquids
\end{title}
\author{D.~Coffey$^a$ and K. S. Bedell$^b$}
\begin{instit}
$^a$ Center for Materials Science, MS-K765, Los Alamos National Laboratory,
Los Alamos, NM 87545
\end{instit}
\begin{instit}
$^b$ Theoretical Division, MS-B262, Los Alamos National Laboratory,
Los Alamos, NM 87545
\end{instit}
\begin{abstract}
We calculate the entropy of a two-dimensional Fermi Liquid(FL)
using a model with a contact
interaction between fermions.   We find that there are $T^2$ contributions to
the entropy from interactions separate from those due to the collective modes.
These $T^2$ contributions arise from non-analytic
corrections to the real part of the self-energy which may be calculated from
the leading log dependence of the imaginary part of the self-energy
through the Kramers-Kronig relation.
We find no evidence of a breakdown in Fermi Liquid theory in 2D
and conclude that FL in 2D are similar to 3D FL's.
\end{abstract}
\narrowtext

The unusual nature of the normal state properties of the high temperature
superconductors (HTS) has generated a new
interest in the metallic phase of strongly correlated electronic materials. In
particular, much attention has
been focused on the existence\cite{EngRan,FHN,SerHes}
or non-existence\cite{PWA1,Varma,ZimBed,SchRin,Stamp} of a Fermi liquid
phase for these systems in
two dimensions (2D).
This controversy as to the existence of 2D Fermi liquid(FL)
is motivated by the difficulty of fitting some experimental
data on the HTS materials with conventional
FL expressions and also by the property of one dimensional systems that the
ground state of a system of interacting fermions
is a Luttinger Liquid (LL) rather than a FL.
In particular this has lead to the development of the
marginal Fermi Liquid (MFL)
phenomenology\cite{Varma,SchRin} which has
been used extensively to fit data.\cite{Bat1}
However there is no microscopic calculation  as yet which leads to a
 an MFL ground state.

The stability of a FL ground state
has been studied extensively in the dilute limit and for weak
coupling\cite{SerHes} in 2D. In the dilute
limit it is possible to show that the particle-particle channel diagrams
contribute in leading order. The
presence of a two hole bound state in this channel lead to speculations that
this was a possible source of the
breakdown of the Fermi liquid phase.\cite{PWA2} It appears, though, that this
bound
state only gives rise to
higher order corrections to the properties of the FL. In weak coupling
away from half filling this
stability of the FL phase of the 2D Hubbard model was also observed in
the propagator renormalized
fluctuation exchange approximation of Serene and Hess.\cite{SerHes} In this
approach all
of the known instabilities,
superconductivity, spin and charge density waves, and the two hole bound state
could occur. No evidence for a
breakdown of the FL phase was observed at quarter filling.

{}From these studies we see that the FL phase of the Hubbard model is
stable against particle-hole or
particle-particle fluctuations away from half-filling. From the two-hole bound
state it was shown\cite{EngRan} that
this contributed a term of the order $|\epsilon|^{5/2}$ to the imaginary part
of
the self-energy, $\Sigma$
(p,$\epsilon$), and from Kramers-Kronig a similar term is found in Re
$\Sigma$(p,$\epsilon$). The two hole
bound states are predominantly short wavelength fluctuations, in this Letter we
investigate the long
wavelength fluctuations. We find that they give rise to lower order
non-analytic
corrections
to FL theory than
does the two hole bound state.
In particular, we find that Re
$\delta\Sigma$(p,$\xi_p) \propto $ sgn ($\xi_p)
{\xi_p}^2$ and that this term gives rise to a T$^2$ correction to the
specific heat, C$_V$. What we learn
from our work and that of references 1 and 2 is that the breakdown of FL
theory must be more subtle
than is found in any of the traditional perturbation theory approaches.
The present calculation can not address the issue raised by
Anderson\cite{PWA1,PWA2}
as to the validity of perturbation theory in 2D except to say that there
is no indication of this from perturbation theory itself.

Apart from the question of stability of the Fermi liquid our results for the
corrections to Fermi liquid
theory are most surprising. In 3D the specific heat is known to have a $T^3$
lnT
correction, i.e. C$_V
\simeq \gamma$ T + $\Gamma_{3D}$ T$^3$ lnT. In the 1D system
the breakdown of the FL can be seen already in the 2$^{nd}$ order
perturbation theory where the specific
heat correction is given by $\delta$C$_V$ $\simeq \Gamma_{1D}$ T lnT. (This is
a
clear signal that perturbation
theory does not work since this is more important than the linear term). In the
2D case we find that C$_V \simeq
\gamma$ T + $\Gamma_{2D}$ T$^2$ +... .
One might have expected this to have a T$^2$lnT
correction by studying the 3D
and 1D behavior, this in fact is not the case.

In order to
determine the leading corrections to a 2D FL due to longwavelength
interactions we consider a system of fermions
which interact via a two-body potential as in Eq.(1).
\begin{equation}
H=\sum_{\vec p, \sigma} {{|\vec p|^2} \over {2m}}
c^\dagger_{\vec p,\sigma}c_{\vec p,\sigma}
+
 \sum_{\vec p,\vec q,\alpha, \beta, \gamma, \delta}
V_{\alpha,\beta,\gamma,\delta}(\vec q)
 c^\dagger_{\vec p,\alpha}c_{\vec p ',\beta}
c^\dagger_{\vec p ' -\vec q,\gamma}c_{\vec p +\vec q,\delta}.
\end{equation}
Expanding in the particle-hole channel the effect of the interaction may
be considered as coming from two independent channels, the symmetric(s)
and the antisymmetric(a) channels corresponding to no spin exchanged
and to spin 1 exchanged.
Using the paramagnon model first introduced by Doniach and
Engelsberg\cite{DonEng},
for the $\vec q$ dependence of the interaction,
the value of the interaction in the symmetric channel,$V_s$, is -I/2
and in the antisymmetric channel, $V_a$, is $2I$, where
$I>0$.\cite{BrinkEng}
I is the the strength of the
interaction and multiplied by the density of states of one spin is the
paramagnon parameter, $\bar{I}$.  The interaction is cutoff at $|\vec q|=q_c$.
In order to compare the properties of FLs in 2D and 3D
we first calculate the single-particle self-energy
to second order in perturbation theory at zero temperature.
For a 3D FL Blaizot and Friman\cite{BlaFri} found
\begin{equation}
Im\Sigma(\vec p, \epsilon) = {{\pi\bar{I}^2}\over{8{v_F}^2{p_F}^2}}
sgn(\epsilon)([q_cv_F -|\xi_p|]{\epsilon}^2 +
{{1}\over{3}}{|\epsilon|}^3 + ....),
\end{equation}
where ${\xi_p}=(p-{p_F}){v_F}$, $p_F$ is the Fermi momentum,
 and ${v_F}={{p_F}/{m}}$
is the Fermi velocity.
The real part of $\Sigma (\vec p,\epsilon)$ is determined from the
Kramers-Kronig relation,
\begin{equation}
Re\Sigma(\vec p, \epsilon ) = {{1}\over{\pi}}P\int^{\infty}_{-\infty}
d\zeta
{{Im\Sigma(\vec p, \zeta)}\over{\epsilon-\zeta}}
\end{equation}
and is given by
\begin{equation}
Re\Sigma(\vec p, \xi_p) = A_{3D}{\xi_p} +B_{3D}{\xi_p}^3\ln|{\xi_p}|+..
\end{equation}
The ${\xi_p}^3 ln|\xi_p|$ term comes from the
$|\epsilon|^3$ term in Im$\Sigma (\vec p,\epsilon)$
and so it is determined by the long-wavelength scattering.
In fact it has been shown by Moriya\cite{Mor} that no ${\xi_p}^3$ln
$|\xi_p|$ terms come from finite $\vec q$ scattering.
For a 2D FL we find
\FL
\begin{eqnarray}
Im\Sigma (\vec p, \epsilon )=sgn(\epsilon )
{{4 \bar{I}^2}
\over {2{\pi}N(0)v_F^2}}
\Biggl[({\xi_p}^2+2{\xi_p}(\epsilon-{\xi_p})) ln(max[{\xi_p}, |\epsilon|])
\Theta({q_c}{v_F}-|\epsilon|)
\\ \nonumber
+({\epsilon} - {\xi_p})^2
ln (|{\epsilon} - {\xi_p} |)
\Theta({q_c}{v_F}-|\epsilon -{\xi_p}|) + ..\Biggr]       .
\end{eqnarray}
for ${\xi_p} > 0$.
This reduces to the well-known results of Hodges et al.\cite{HSW}
and Bloom\cite{Blo} for Im$\Sigma(\vec p,\epsilon = \xi_p)$.
Using the Kramers-Kronig relation again, Re$\Sigma(\vec p, \xi_p)$ is
\begin{equation}
Re\Sigma(\vec p, {\xi_p}) = A_{2D}{\xi_p} +B_{2D}sgn({\xi_p}){\xi_p}^2 +.......
\end{equation}
Only the first term in  Im$\Sigma(\vec p, \epsilon)$ contributes to
 Re$\Sigma(\vec p, \xi_p)$.
The linear terms in Re$\Sigma(\vec p, \xi_p)$ for 2D and 3D
are effective mass enhancements,
which come from all $\vec q$'s and depend on the $\vec q$ dependence
of the interaction.  Here we are concerned with the corrections
to the effective mass enhancement terms.
As in the case of $B_{3D}$, $B_{2D}$ is determined by long wavelength
scattering.
Comparing the corrections to the effective mass terms in
Re$\Sigma (\vec p , {\xi_p})$ in 2D and 3D,
one sees that in 2D the correction is non-analytic
and comes from the leading $\epsilon$
dependence of Im$\Sigma (p,\epsilon )$ whereas
in 3D the correction is analytic but comes from non-analytic terms in
Im$\Sigma (p,\epsilon )$.
This difference between the leading corrections to the linear
$\xi_p$ dependence in 2D and 3D is due solely to the different phase space.
This may be seen by calculating the contribution to the spectrum
in 2D using the equation,
\begin{equation}
\Delta \epsilon_p = \sum_{|\vec q | < q_c}
[1-2f_{\vec p +\vec q}] {(\hat{p} . \hat{q} )^2} {V^{(2)}},
\end{equation}
where $V^{(2)}$ is the coefficient of the $(\hat{p} . \hat{q} )^2$
term in the effective quasiparticle interaction.  In 3D this gives the
$\xi_p^3ln|\xi_p|$ dependence.\cite{BPet}
In contrast to 1D, where the analogous calculation already shows that
FL theory has broken down, there is no indication of a breakdown of FL
theory in 2D to this order in perturbation theory.  We now consider the
thermodynamics of a 2D FL and compare the results with 3D.

Using the RPA approximation the change in the thermodynamical
potential due to interactions
in Eq.(1) is given by
\begin{equation}
{\Delta \Omega }={k_B T} \sum_{\vec q ,\omega_n}
\Biggl[
{{3}\over{2}}(ln[1-I \chi (\vec q,\omega_n)] + I \chi (\vec q, \omega_n) )
\\ \nonumber
+{{1}\over{2}} (ln[1+I\chi (\vec q,\omega_n)] -
 I \chi (\vec q, \omega_n) ) \Biggr]
\end{equation}
where
\begin{equation}
\chi(\vec q,\omega ) = 2\sum_{\vec p} {{f_{\vec p + \vec q}-f_{\vec p}}
\over{\omega - (\epsilon_{\vec p + \vec q} - \epsilon_{\vec p})}},
\end{equation}
$f_{\vec p }$ is the Fermi Dirac distribution function,
and $\omega_n=2\pi $(n+1)T are Matsubara frequencies.
When analytically continued to the real $\omega $ axis $\Delta \Omega $ can be
easily broken up into a quasiparticle contribution, $\Delta \Omega_{qp}$
, and a contribution from collective modes, $\Delta \Omega_{coll. modes}$.
First we consider $\Delta \Omega_{qp}$ which is given by
\begin{equation}
\Delta \Omega_{qp} = \sum_{|\vec q|<q_c} \int^{\infty}_0
{{d\omega} \over {\pi}}
    n_B(\omega )(F(\vec q,\omega ) +I\chi^{\prime\prime}(\vec q, \omega ))
\end{equation}
\begin{equation}
F(\vec q,\omega )={{3}\over{2}}
\tan^{-1}[{{-I\chi^{\prime\prime}(\vec q, \omega )} \over
 {1 - I\chi^\prime(\vec q,\omega )}}]
+ {{1}\over{2}}
\tan^{-1}[{{I\chi^{\prime\prime}(\vec q, \omega )} \over
 {1 + I\chi^\prime(\vec q,\omega )}}],
\end{equation}
\begin{equation}
\chi(\vec q, \omega ) = \chi^\prime(\vec q, \omega ) +
 \imath \chi^{\prime\prime}(\vec q, \omega ),
\end{equation}
and $n_B(\omega )$ is the Bose distribution function.
{}From this the change in the entropy is
\begin{eqnarray}
{\Delta S_{qp} } &=& -\Biggl[
{ {\partial {\Delta \Omega_{qp}} } \over {\partial T} }
\Biggr]
_{\mu}
\\ \nonumber &=&\sum_{|\vec q|<q_c} \int^\infty_0
{{d\omega}\over{\pi}}
({{\partial {n_B(\omega)}}\over{\partial T}}\Biggr|_{\mu}
F(\vec q,\omega) +
n_B(\omega){{\partial {F(\vec q, \omega)}}\over{\partial T}}\Biggr|_{\mu}  +
I{{\partial{( n_B(\omega) \chi^{\prime\prime}(\vec q, \omega))}}
\over{\partial T}}\Biggr|_{\mu})
\nonumber
\end{eqnarray}
The two terms on the right of Eq.(13) involve the
temperature dependence of $\chi (\vec q, \omega)$
which is weak when $\mu $ is kept constant.

Calculating $\Delta S_{qp}$ for 2D one finds
\begin{equation}
\Delta S_{qp} =
       \gamma_{2D}\prime T + \Gamma_{2D} T^2 + O(T^3)
\end{equation}
\begin{eqnarray}
\gamma_{2D}\prime = {{\pi}\over{6T_F}} (A_s+A_a) {{q_c}\over{p_F}}
\nonumber\\
\Gamma_{2D}= {{\pi n}\over {4 T^2_F}} \sum_{\lambda} \nu_{\lambda}
(A_{\lambda}+\int^1_0 du f_{\lambda}(u))
\end{eqnarray}
where
\begin{equation}
f_{\lambda}(u)=
{ { [ A_{\lambda}u-tan^{-1}({{A_{\lambda}u}\over{1-u^2}}) ] }
\over {u^3} },
\nonumber
\end{equation}
where $\lambda$=s or a, $\nu_s$=1, $\nu_a$=3, and
\begin{equation}
A_s={{\bar{I}}\over{1+\bar{I}}},     A_a={{-\bar{I}}\over{1-\bar{I}}}
{}.
\end{equation}
$A_s$ and $A_a$ are the scattering amplitudes in the symmetric(density) and
antisymmetric(spin) channels,
n is the density of particles, and $T_F = {v_F}{p_F}/2$.
The $T^2$ term in $\Delta S_{qp}$ comes from the non-analytic term in
Re$\Sigma (\vec p, \xi_p)$ as may be seen from the
following argument.\cite{Pet2}
Consider the entropy of a FL whose spectrum is given by
\begin{equation}
{\epsilon_p} = {\xi_p} + \Delta {\epsilon_p}
\end{equation}
where $\Delta {\epsilon_p}$ arises from interactions.
Substituting this spectrum into the expression for the entropy of a
non-interacting Fermi liquid and expanding to linear order in
$\Delta {\epsilon_p}$
one finds
\begin{equation}
\Delta S = \sum_{\vec p} {{\xi_p} \over {T}} {{\Delta \epsilon_p} \over {T}}
{{\partial f(\epsilon)} \over {\partial \epsilon}}
\Biggl|_{\xi_p}
\end{equation}

Assuming that $\Delta {\epsilon_p}$ can be
expanded in a power series in $\xi_p$
one finds
\begin{equation}
\Delta S = N(0)k_B \int_{-\infty}^{\infty} {{\xi d\xi}\over{T^2}}
{{1} \over {4 \cosh^2{ {\xi}\over{2T} } } }
\sum_n \alpha_n \xi^n
\end{equation}
One sees that only odd n terms contribute to $\Delta S$ and that they lead
to series of odd powers of T.
The presence of the $T^2$ in $\Delta S_{qp}$ clearly arises from the
non-analytic nature of the correction to the spectrum in Eq.(6).
The terms of O($T^3$) and higher are a sum of odd powers of temperature.

Carrying out the calculations in 3D one finds that
\begin{equation}
\Delta S_{qp} =
       \gamma_{3D}\prime T + \Gamma_{3D} T^3 ln T + O({T^3})
\end{equation}
Another difference between 2D and 3D is that
$\Gamma_{2D}$ depends on the scattering amplitudes to all orders whereas
$\Gamma_{3D}$ involves only the second and third powers of the scattering
amplitudes.

The collective mode contribution in 2D is
\begin{equation}
\Delta S_{coll.mode}=\Gamma^{\prime\prime} T^2 + O(T^4)
\end{equation}
where
\begin{equation}
\Gamma^{\prime\prime} = {{1}\over{2\pi}} ({{T}\over{c}})^2
\end{equation}
and c is the velocity of the collective mode given by
\begin{equation}
c= {v_F}{{1+\bar{I}} \over {\sqrt{1+2 \bar{I}}}}
\end{equation}
The collective mode spectrum does not contain any log dependence on
$|\vec q|$ and so there is no $\ln T$ contribution to $\Delta S_{coll.mode}$.
In 3D $\Delta S_{coll.mode}\sim T^3$ and does not
contribute to the $T^3lnT$ corrections except to change the
cutoff of the logarithmic temperature dependence.

Collecting $\Delta S_{qp}$ and $\Delta S_{coll.mode}$ together
one sees that $\Delta S$ is a power series in temperature in 2D.
In particular there are no $\ln T$ terms in $\Delta S$ which
implies that,
at least to leading order,
quasiparticle damping effects do not contribute
to thermodynamic properties in this approximation
where the propagators are unrenormalised.
The effect of finite quasiparticle lifetimes on the entropy
may be estimated with Eq.(23)
of ref.\cite{CPet1} and is given by
\begin{equation}
\Delta S_{damp} = \sum_{\vec p} \int_{-\infty}^{\infty}d\epsilon
{{\partial f(\epsilon)} \over {\partial T}}
G({\lambda(\vec p, \epsilon)}),
\end{equation}
where
\begin{equation}
G({\lambda(\vec p, \epsilon)}) =
{{\lambda(\vec p, \epsilon)} \over {1+{\lambda(\vec p, \epsilon)}^2}} -
tan^{-1} ({\lambda(\vec p, \epsilon)}),
\end{equation}
and
\begin{equation}
\lambda(\vec p, \epsilon) = -{{Im \Sigma(\vec p, \epsilon)} \over
{Re\Sigma(\vec p, \epsilon)}}.
\end{equation}
The functional form of the integrand in Eq. (25) is very complicated
and analytic evaluation is intractable.  So we content ourselves with an
estimate.  $G({\lambda(\vec p, \epsilon)}$
is a smooth function of $\lambda(\vec p, \epsilon)$
which goes as
${\lambda(\vec p, \epsilon)}^3$  for small $\lambda(\vec p, \epsilon)$
and is a constant  for large $\lambda(\vec p, \epsilon)$.
In order to get an estimate of $\Delta S_{damp}$ we
assume that
$G \sim {\lambda(\vec p, \epsilon)}^3$ for all values of $\epsilon$.
Since there is no contribution to the integral
for large values of $|\lambda(\vec p, \epsilon)|$,
this is clearly an overestimate.
With these approximations one finds $\Delta S_{damp} \sim T^7ln^3T$
which is
higher order in temperature than the
$T^2$ corrections found above.
Lifetime effects lead to higher order effects in the thermodynamics than
$T^2$ so that they are much less important in thermodynamics than a
calculation of $\Sigma(\vec p$, $\epsilon$) would suggest.

Since we have used a
particle-hole expansion in the symmetric and antisymmetric channels our
results may be easily extended to Landau's Fermi liquid theory by considering
Eq.(1) to describe quasiparticles with an effective mass interacting via
an effective interaction $f(\vec p,\vec p^\prime)$ which can be decomposed
into two channels, $f_s(\vec p,\vec p^\prime)$ and $f_a(\vec p,\vec p^\prime)$.
This effective interaction is a long wavelength limit of the
particle-hole irreducible four-point vertex and so describes long
wavelength properties.
This leads to somewhat more complicated expressions when $V_s$ and $V_a$ are
substituted for by $f_s(\vec p,\vec p^\prime)$
and $f_a(\vec p,\vec p^\prime)$.
The Landau functions  are functions of the variable
$s = {{\omega} \over {qv_F}} $
and may be expressed as coefficients in a series of Legendre
polynomials in which $s$ is the argument. These coefficients are the Landau
parameters.

The present calculation indicates that a 2D FL is very
similar to the 3D case and that any breakdown of the FL in 2D
has to arise from
effects which are more subtle than those which give the leading corrections to
FL theory in 3D.
The logarithmic dependence in  Im$\Sigma$($\vec p$, $\epsilon$)
allow us to keep track of the contribution to the thermodynamic properties
from lifetime effects.
We find that in spite of the ${\xi_p}^2 ln{\xi_p}$ dependence of the
relaxation time in 2D their contributions are higher order in T than
the contributions to the quasiparticle spectrum.
\bigskip

The authors would like to thank C. J. Pethick for several important
contributions to this work.
DC acknowledges the support of the Center for Materials
Science through the Program in Correlated Electron Theory.
This work was supported by the US Department of
Energy.



\begin{references}

\bibitem[1]{EngRan} J. R. Engelbrecht and M. Randeria,
Phys. Rev. Lett. {\bf 65}, 1032
(1990)  and Phys. Rev. B {\bf 45}, 12419 (1992).

\bibitem[2]{FHN} H. Fukuyama, Y. Hasegawa and O. Narikiyo, J. Phys. Soc. Jap.
{\bf 60}, 2013 (1991).

\bibitem[3]{SerHes}
J. W. Serene  and D. W. Hess, Phys. Rev. B {\bf 44}, 3391 (1991).

\bibitem[4]{PWA1}
P. W. Anderson, Phys. Rev. Lett. {\bf 64}, 1839 (1990).

\bibitem[5]{Varma}
C. M. Varma, P. Littlewood, S. Schmitt-Rink, E. Abrahams and A. E. Ruckenstein,
Phys. Rev. Lett. {\bf 63}, 1996 (1989).

\bibitem[6]{ZimBed}
G. Zimanyi and K. S. Bedell, Phys. Rev. Lett. {\bf 66}, 228 (1991).

\bibitem[7]{SchRin}
S. Schmitt-Rink, C. M. Varma and A. E. Ruckenstein, Phys. Rev. Lett. {\bf 63},
445 (1989).

\bibitem[8]{Stamp}
P. C. Stamp, Phys. Rev. Lett. {\bf 68}, 2180 and 3938 (1992).

\bibitem[9]{Bat1}
B. Batlogg in High Temperature Superconductivity: Proceedings of the Los Alamos
Symposium 1989, edited by K. S. Bedell, D. Coffey, D. E. Meltzer,
D. Pines and J. R. Schrieffer,
(Addison-Wesley, Redwood City, California, 1990), page 37.

\bibitem[10]{PWA2}
P. W. Anderson, Phys. Rev. Lett. {\bf 65}, 2306 (1990).

\bibitem[11] {DonEng} S. Doniach and S. Engelsberg, Phys. Rev. Lett., {\bf 17},
750 (1966).

\bibitem[12]{BrinkEng} W. F. Brinkman and S. Engelsberg, Phys. Rev. {\bf 169},
417 (1968).

\bibitem[13]{BlaFri} J. P. Blaizot and B. L. Friman, Nucl. Phys. {\bf A372},
69 (1981)
{}.

\bibitem[14]{Mor} T. Moriya, Phys. Rev. Lett. {\bf 24}, 1433 (1970)
and T. Moriya and T. Kato J. Phys. Jap. Soc. {\bf 31}, 1016 (1971).

\bibitem[15]{HSW} C. Hodges, H. Smith and J.W. Wilkins,
Phys. Rev. {\bf 4}, 302 (1971).

\bibitem[16]{Blo} P. Bloom, Phys. Rev. B {\bf 12}, 125 (1975).

\bibitem[17]{BPet} G. Baym and C. J. Pethick,
in "The Physics of Liquid and Solid
Helium", edited by K. H. Bennemann and J. B. Ketterson
(Wiley, New York, 1978) Part II, Eq.(1.4.69), page  102.

\bibitem[18]{Pet2} C. J. Pethick, (private communication).

\bibitem[19]{CPet1} G. M. Carneiro and C. J. Pethick, Phys. Rev. B {\bf 11},
1106
(1977).

\end{references}
\end{document}